\newfont {\xx} {cmti10}

\newcommand{\bea}{\begin{eqnarray}}
\newcommand{\eea}{\end{eqnarray}}
\documentstyle[epsfig,epsf,12pt]{article}
\begin{document}
\pagestyle{empty}
\begin{flushright}
{CERN-TH/96-101} \\
{OUTP-96 20P} \\
\end{flushright}
\vspace*{5mm}
\begin{center}
{\bf CAN ONE PROBE THE STRUCTURE FUNCTION OF THE POMERON?} \\
\vspace*{1cm}
{\bf John Ellis}\\
\vspace{0.3cm}
Theoretical Physics Division, CERN \\
1211 Geneva 23, Switzerland\\
and \\
{\bf Graham G. Ross} \\
\vspace{0.3cm}
Theoretical Physics, 1 Keble Rd.\\
Oxford OX1 3NP, United Kingdom\\
\vspace*{2cm}
{\bf ABSTRACT} \\ \end{center}
\vspace*{5mm}
\noindent
We discuss whether the diffractive structure functions defined by current
experiments at HERA are indeed probing the partonic structure function of
the pomeron. We observe that the {\it pseudorapidity} cuts commonly employed
require that the struck parton in the pomeron be far off mass shell in
sizeable regions of parameter space. As a result an interpretation in terms
of constituent partons within the pomeron is inadequate. One may
nevertheless use a partonic description for the {\it amplitude} for virtual
photon-pomeron scattering to compute a diffractive structure function for
pseudorapidity gap events. The resulting form may have significant
scaling violation.

\vspace*{5cm}

\begin{flushleft}
CERN-TH/96-101\\
OUTP-96 20P\\
April 1996
\end{flushleft}
\vfill\eject

\setcounter{page}{1}
\pagestyle{plain}




\section{Introduction}

There has been intense interest in the interpretation of the large {\it %
pseudorapidity} gap events observed at HERA in deep-inelastic scattering
processes \cite{Z1,Z2,Z3,H1}. Such events must involve the exchange of a
colourless state between the proton and the virtual photon, and, at high
energies, the diffractive component corresponding to pomeron exchange will
be dominant. Assuming that this component is responsible for the observed
events, the data lead to the determination of a diffractive structure
function. It has been suggested by Ingelman and Schlein \cite{is} that such
measurements should allow the distribution of quark and gluon partons within
the pomeron to be determined, which would clearly be a very interesting
possibility \cite{dl}.

In this letter we consider the extent to which the current experiments at
HERA address the Ingelman-Schlein proposal. We argue that the kinematics of
experiments that use a strong cut in {\it pseudo}rapidity (i.e. $%
\eta=-ln(tan(\theta_{lab}/2))$) to define a diffractive structure function
do not admit a simple interpretation in terms of the partonic structure of
the pomeron, because the struck quark or gluon cannot always be close to its
 mass
shell. However, we argue that, even with such a pseudorapidity cut, there is
a valid description in which an off-shell parton emanates from the virtual
photon and scatters diffractively off the proton\footnote{%
A similar proposal has been made in different terms by Bjorken \cite{bj}. For a
 related discussion of the interpretation of the rapidity gap events in terms of
 a simple gluon structure function, see \cite{buch}.}
This should be evaluated by computing the full photon-pomeron scattering
amplitude, thus retaining the coherence effects involved in having the
struck parton far off mass shell. We present such a calculation, and show
that the resulting cross section can be written in a factorised form, in
which the diffractive structure function exhibits a modification of the
usual scaling behaviour which are characteristic for the process.

The various experimental cuts used have differing sensitivity to the virtuality
 of the struck parton. For example the ZEUS Collaboration has recently published
 a new extraction of a diffractive structure function which does not involve a
 pseudorapidity cut,
but makes an event selection based on the invariant mass of the hadronic
system produced in association with unseen remnants of the proton \cite{Z3}.
Thus these experiments will have a component of varying importance which may be
 interpreted  in terms of parton
distributions within the pomeron, but all will also involve a significant
 component involving far-off-mass-shell partons.

\section{Kinematics of Diffractive Deep-Inelastic Scattering}

The experimental results \cite{Z1,Z2,Z3,H1} are usually presented in a form
analogous to that of the total deep-inelastic scattering cross section,
namely

\begin{equation}
\frac{d^3 \sigma_{diff}}{d\beta dQ^2 dx_{{\cal P}}}=\frac{2\pi\alpha^2}{%
\beta Q^4}(1+(1-y)^2)F_2^{D(3)}(\beta ,Q^2, x_{{\cal P}})
\end{equation}
where the contribution of $F_L$ is neglected. The effect of neglecting $F_L$
corresponds to a relative reduction of the cross section at small $x_{{\cal P%
}}$ (high $W^2$) which is always $<17$\% \cite{Z1,Z2,Z3,H1}, and therefore
smaller than the typical present measurement uncertainties $(\simeq 20$\%).

\begin{figure}[htp]
\hglue 5.5cm
\epsfig{figure=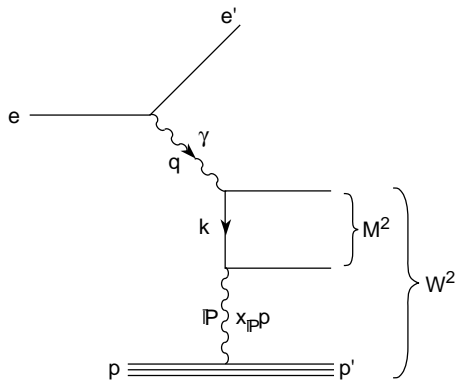,width=6.0cm}
\caption[]{Quark-parton graph contributing to the large-rapidity-gap
diffractive scattering amplitude. }
\label{fig:1}
\end{figure}

The variables $Q^2$ and $y$ have the definitions usual for deep-inelastic
scattering. The variables $x_{{\cal P}}$ and $\beta$ are defined as
\begin{eqnarray}
x_{{\cal P}}=\frac{(P-P^{\prime}).q}{P.q} \simeq \frac {M^2 +Q^2}{W^2+Q^2}
\nonumber \\
\\
\beta = \frac{Q^2}{2(P-P^{\prime}).q} \simeq \frac{Q^2}{M^2+Q^2}
\end{eqnarray}
where $W^2$ and $M^2$ are the total hadronic invariant mass squared and the
mass squared of the hadrons excluding the proton remnants, respectively. In
the framework of the underlying quark parton diagram of Fig.~\ref{fig:1},
these would normally be interpreted as the momentum fraction of the pomeron
within the proton and the momentum fraction of the struck quark within the
pomeron, respectively.

We argue that the kinematics of the initial HERA experiments are such that
one cannot in fact interpret the graph of Fig.~\ref{fig:1} in terms of a
conventional parton density within the pomeron. The reason is that, if one
wants a probabilistic interpretation of Fig \ref{fig:1}, with the cross
section factorized as the product of an elementary subprocess cross section
with the probability to find a parton within the pomeron viewed in the
direct channel, the parton should be near its mass shell. On the other hand,
in order to interpret the graph as a diffractive process, the sub-energy $%
(k+p)^2 \equiv s_{diff}$ should be large enough for the leading singularity
in the cross channel, namely the pomeron, to be dominant, giving rise to an
amplitude behaving as $s_{diff}^{\alpha_{{\cal P}}-1}$, where $\alpha_{{\cal %
P}}$ is the intercept of the pomeron Regge trajectory. This leads to the $x_{%
{\cal P}}^{-(2\alpha_{{\cal P}}-1)}$ dependence of the cross section which
is indicated by experiment. The trouble with the partonic interpretation of
these diffractive events arises because it requires that the parton has a
longitudinal momentum equal to some fraction of that of the pomeron, $\beta
P_{{\cal P}}$, which in turn carries only a small fraction of the proton
momentum, $P_{{\cal P}}=x_{{\cal P}} P$. Thus, if the partonic
interpretation applies, $k\approx\beta P_{{\cal P}} =\beta x_{{\cal P}} P$,
and so $s_{diff}=(k+P)^2$ is small, in potential conflict with the
requirement that the process be diffractive. Indeed, as discussed below, one
finds that, for the strongest pseudo-rapidity cuts for a range of $x_{{\cal P%
}},\;\beta$ and $Q^2$, the dominant part of the cross section comes from $%
k^2 \sim Q^2$, rather than $k^2$ small and near mass shell. Thus the
interpretation of Fig.~\ref{fig:1} is not the standard one for
deep-inelastic scattering off a hadronic target, but one in which the struck
parton is far off shell. This means that the graph of Fig.~\ref{fig:1}
cannot be calculated as the product of two separate cross sections $%
\sigma_{q/{\cal P}}$.$\sigma_{\gamma q \rightarrow X}$, but must instead be
considered as a complete amplitude $A_ {\gamma {\cal P \rightarrow X}}$.
This also means that the interpretation of the analysis as a determination
of the parton distribution within the pomeron needs re-evaluation.

\section{Parton Virtuality in Diffractive Scattering}

In order to quantify our claim that, at least for some of the experimental
cuts, the dominant part of the diffractive cross section comes from a struck
parton far off shell, let us consider Fig.~\ref{fig:1} again in more detail.
The condition that the final state quark is on-shell implies
\begin{equation}
k^2= -2 x_{{\cal P}} k.p=-x_{{\cal P}} s_{diff}  \label{onshell}
\end{equation}
If this process is to be dominated by pomeron  exchange, $s_{diff}$ should
be large. However,  since $x_{{\cal P}}$ can be very small,  this constraint
does not by itself require that  the struck parton be far off-mass-shell.
Working in the photon-pomeron centre of mass frame,  one readily determines
that
\begin{equation}
k^2 =\frac{M_X^2 + Q^2}{2} (1-\cos\theta_{cm})  \label{cmangle}
\end{equation}
where $\theta_{cm}$ is the centre-of-mass angle between  the proton
direction and the final-state quark. Apart from the region $\cos
\theta_{cm}\simeq 1$, the struck quark is clearly off mass shell.

\begin{figure}[htp]
\hglue 6.5cm
\epsfig{figure=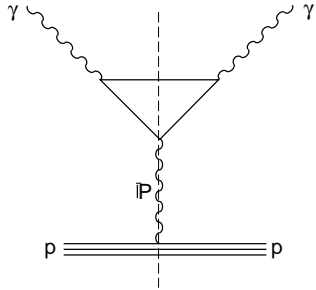,width=4.1cm}
\caption[]{Quark-parton graph contributing to the inclusive diffractive cross
section.  }
\label{fig:2}

\end{figure}

The Ingelman-Schlein proposal applied to the quark constituents of the
pomeron has been developed by Donnachie and Landshoff \cite{dl}. They argue
that this process is indeed dominated by the region $\cos \theta_{cm }\simeq
1$, and that the quark-pomeron coupling has a form factor that falls rapidly
for large $k^2$, so that $k^2 \le1 GeV^2$ gives the only significant
contribution. They reach this conclusion by considering the inclusive
diffractive cross section, i.e., including events which have no large
rapidity gap. These events are expected to be given by the imaginary part of
the graph of Fig.~\ref{fig:2}.
Donnachie and Landshoff propose a pomeron-quark coupling
of the Wu-Yang form
\begin{equation}
\beta_0 f(k^2_1)f(k^2_2)\bar{q}\gamma_{\mu}q{\cal P}
\label{eq:donland}
\end{equation}
where $k_1$ and $k_2$ are the virtualities of the initial and final quark,
and $\beta_0$ is a coupling with the dimension of an inverse mass. With the
form factors $f(k^2)$ omitted, calculation of the contribution of Fig.~\ref
{fig:2} leads to a structure function proportional to $\beta_0^2 Q^2$, which
clearly does not scale. Donnachie and Landshoff therefore argue that the
form factors $f(k^2)$ should be included, and choose them
phenomenologically, such that the cut off of the loop integral occurs at a
hadronic scale $\Lambda$ with $\Lambda=0(1GeV)$. This leads to a structure
function proportional to $\beta_0^2 \Lambda^2$, which scales, and is
consistent with observation. With this motivation, they apply the same
vertex to the calculation of Fig.~\ref{fig:1}. In this case, the form factor
keeps $k^2$ close to the mass shell, so that the process can indeed be
interpreted in the parton-model sense as a convolution of the photon-quark
scattering cross section with the probability of finding a quark within the
pomeron.

In our opinion, this is not the correct conclusion, for two reasons. The
first is purely phenomenological, and relates to the experimental cuts
imposed to define the diffractive events. Consider again the graph of Fig.~%
\ref{fig:1}, and consider the implication of imposing a pseudorapidity cut
on the data, which in turn requires $\cos \theta \geq \cos \theta
_{cm}^{\min }$. Clearly, this implies
\begin{equation}
k^2\geq k_{min}^2=\frac{M_X^2+Q^2}2(1-\cos \theta _{cm}^{\min })
\label{offshell}
\end{equation}
and forces the struck quark far off shell if $\cos \theta _{cm}^{\min }<1$.
A variety of experimental cuts have been employed in different experimental
papers. In the first ZEUS paper, $\eta _{min}=1.5$ \cite{Z1}, while in the
second ZEUS paper $\eta _{min}=2.5$ \cite{Z2}, and the $H1$ collaboration
uses $\eta _{min}=3.2$ \cite{H1}. In all cases, the events selected have an $%
x_{{\cal P}}$ dependence consistent with their interpretation as diffractive
scattering events due to pomeron exchange with a trajectory $\alpha _{{\cal P%
}}\geq 1$.

In order to translate these cuts into a value for $\cos \theta _{cm}^{\min }$%
, and hence determine the constraint on the off-shell mass of the struck
quark, we first note that $\eta _{min}$ refers to the minimum pseudorapidity
of all calorimeter clusters in an event, where a cluster is defined as an
isolated set of adjacent cells with summed energy above $400MeV$. The
interpretation of $\eta _{min}$ for the graph of Fig.~\ref{fig:1} clearly
requires some information how the jet associated with a final-state quark or
antiquark develops. In jet studies at ZEUS, the cone radius $R={(\Delta \phi
^2+\Delta \eta ^2)}^{1/2}$ was set to one unit and gave results consistent
with QCD expectations. A jet associated with a primary parton will spread in
rapidity and, if we use the ZEUS algorithm, we may expect the spread to be
of order 1/2 to 1 unit of rapidity. The numbers in Table 1 are derived using
the smaller value of $1/2$ for the spread in rapidity. A more
detailed calculation requires a full Monte Carlo simulation using the
detailed experimental cuts and will be sensitive to the details of the model
used for the jet development.

\begin{table}[t]
\centering
\begin{tabular}{|ccccc|}
\hline
&  &  &  &  \\
Q$_{}^2$ & $\beta $ & x$_P^{}$ & -k$_{\min }^2\;GeV^2$ & -k$_{\min }^2\;GeV^2$
 \\
&  &  & $(\eta _{\min }=1.5)$ & $(\eta _{\min }=2.2)$ \\
&  &  &  &  \\ \hline
&  &  &  &  \\
10 & 0.175 & .0032 & 3.1 & 0.4 \\
&  & .0050 & 7.5 & 1.0 \\
& 0.375 & .0020 & 0.9& 0.12 \\
&  & .0032 & 2.3 & 0.3 \\
& 0.65 & .0013 & 0.2  & 0.03 \\
&  & .0020 & 0.5 & 0.07 \\
28 & 0.175 & .005 & 7.5 & 1.0 \\
&  & .0079 & 18.7 &2.5 \\
& 0.375 & .002 & 0.9 & 0.1 \\
&  & .0079 & 14.2 & 1.9 \\
& 0.65 & .0020 & 0.5& 0.07 \\
&  & .005 & 3.2 & 0.4 \\
63 & 0.375 & .005 & 5.7 & 0.8 \\
&  & .0079 & 14.2 & 1.9 \\
& 0.65 & .0032 & 1.3 & 0.2\\
&  & .0079 & 8.0 & 1.1 \\ \hline
\end{tabular}
\label{table:1}
\caption{Minimum virtuality of the struck quark following from a
pseudorapidity cut.}
\end{table}

If the diffractive events observed at HERA are due to the graph of Fig.~\ref
{fig:1}, the lower bound on $k^2$ following from the experimental $\eta
_{min}$ cuts must be satisfied. These apply in the laboratory frame, and the
boost from the laboratory to the centre-of-mass frame depends on the
kinematic variables. In Table~1 we give the bounds $k_{min}^2$ for a range
of these parameters and of $\eta _{min}$ values used in the analysis of the
experimental data\footnote{%
The cuts listed are appropriate to the ZEUS data only. The H1 cuts are
weaker, and correspond to very small $k^2_{min}$.}. The important point to
note is that, over significant ranges of these parameters, $k^2$ is
constrained to be large, far from the hadronic mass scale needed to justify
the interpretation of constituent partons within a pomeron structure
function. It is impressive that the data obtained using these cuts requiring
large $k^2$ have the same diffractive characteristics ($x_{{\cal P}}$
dependence, etc.) as those with weaker cuts. Given this  we do not  see how
one can consistently interpret the rapidity gap events in terms of a
mechanism which requires that $k^2$ be small.

\begin{figure}
\begin{minipage}[t]{.49\textwidth}
\hglue 2.0cm
\epsfig{figure=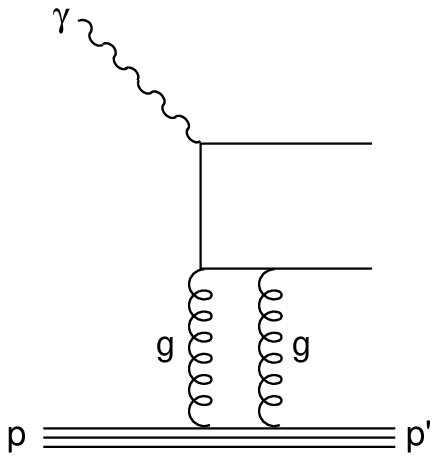,width=3.7cm}
\caption[]{A two-gluon component of Fig.~1 contributing to the
large-rapidity-gap diffractive scattering amplitude.}
\label{fig:3}
\end{minipage}
\hfill
\begin{minipage}[t]{.49\textwidth}
\hglue 1.5cm
\epsfig{figure=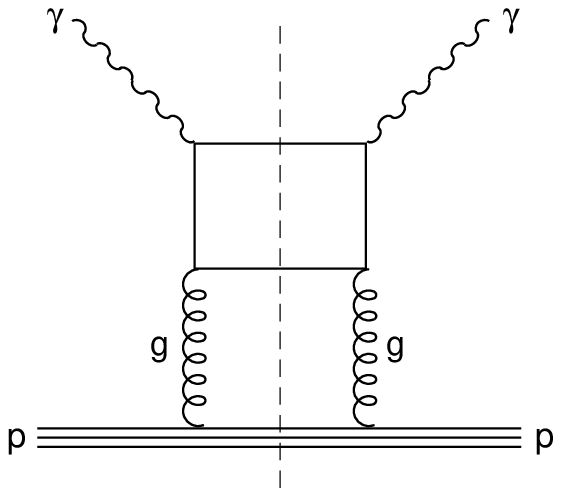,width=4.5cm}
\caption[]{A two-gluon component of Fig.~2 contributing to the inclusive
diffractive cross section. }
\label{fig:4}
\end{minipage}
\end{figure}

The second reason for including high virtuality partons in calculating Fig
 \ref{fig:1} follows because a study of the perturbative (BFKL)
 pomeron \cite{bfkl} suggests that
the form factor $f(k^2)$ does {\it not} cut
off the integral at low $k^2$. Its general behaviour is illustrated
by the two-gluon component, which yields diagrams of the type shown in
Figs.~\ref{fig:3} and \ref{fig:4} for the rapidity-gap and inclusive
processes respectively. In both diagrams the momentum distribution of the
two-gluon component of the pomeron
should be cut off at the hadronic scale $\Lambda$.
As a result, one may see in Fig \ref{fig:3} that there is an additional
 fermion
propagator which will introduce a convergence factor at large quark virtuality,
 $k^2$.
The same is true of the graph in Fig \ref{fig:4}. Evaluating the amplitude
squared of Fig \ref{fig:3} gives an additional term proportional to $\Lambda
^2/k^2$ for large $k^2$,
 when compared to the determination of the amplitude
squared following from Fig.~\ref{fig:1} with a pointlike pomeron-quark
coupling. Evaluating the amplitude squared of Fig \ref{fig:4} also gives a
term proportional to $\Lambda ^2/k^2$ for large $k^2$,
 when compared to that
following from Fig.~\ref{fig:2},
 again with a pointlike pomeron coupling. If
one wishes to interpret this behaviour  in terms of an effective quark-pomeron
coupling we must choose $f(k^2)$ in  eq \ref{eq:donland} to be of the form

\begin{equation}
{f}(k^2)^2=\frac{\Delta ^2}{(k_1^2+k_2^2+\Delta^2)}
\label{eq:donlandr} \\
\end{equation}
This is to be compared with the choice of Donnachie and Landshoff \cite{dl} who
 identify the right hand side with $f(k^2)$. Our form is necessary if we are to
 avoid the problematic strong sensitivity to the pseudorapidity cuts just
 discussed.
At first sight it might seem that the graph of Fig \ref{fig:3} should be
re-interpreted as probing the {\it gluon} component of the pomeron with the
 gluon
constrained to be near its mass shell. However,
this is not the case, because
of the constraint that there should be a rapidity gap between the proton and
the final-state quark or anti-quark. This requires that the exchanged
particle (the pomeron) be a colour singlet,
so that the two-gluon component
is the leading one, i.e.,
 one cannot simply treat the second gluon as a spectator
particle in the final state,
as would be required for the gluon parton interpretation. The
appearance of the $\Lambda ^2/k^2\ $factor is simply a reflection of the fact
that the parton components of the pomeron are made of field components with
dimension $>1$. Of course,
 higher-dimension components will be more
 convergent, so the final form factor need not have the simple power
behaviour shown in (\ref{eq:donlandr}). Note that our approach differs from that
 of \cite{nz,diehl} in that we do not {\it replace} the pomeron contribution
 with a two gluon component because the former is clearly a non-perturbative
 object. Our discussion of the two-gluon component was merely a guide to what it
 is reasonable to expect in the pomeron quark coupling.

\section{Calculation of the Quark Contribution to Large Pseudorapidity Gap
Diffractive Events}
Although, as we have seen, the rapidity gap events measured at HERA do {\it not}
 directly probe the distribution of partons within the pomeron,
 the graph of Fig.
 \ref{fig:1} and the crossed graph may still be relevent provided one drops the
 assumption that the
struck quark is close to its mass shell. In evaluating this graph, we
continue to use the modified Wu-Yang form of the quark-pomeron coupling of
(\ref{eq:donlandr}) discussed above.
The calculation is straightforward, and
leads to the following form:
\begin{eqnarray}
\frac{dF_2^{D(3)}(\beta ,Q^2,x_{{\cal P}})}{dt_{{\cal P}}d\Omega _{q\bar{q}}}%
&=&\beta _0^4[F(t_{{\cal P}})]^2\Lambda ^2(\frac 1{x_{{\cal P}}})^{2\alpha _{%
{\cal P}}(t_{{\cal P}})-1}\nonumber \\
&&((\frac{-t}{\Lambda ^2})^{(\alpha _{{\cal P}}(t_{%
{\cal P}})-1)}f(t)+(\frac{-u}{\Lambda ^2})^{(\alpha _{{\cal P}}(t_{{\cal
 P}%
})-1)}f(u))^2\beta (1-\beta )
\label{eq:f2}
\end{eqnarray}
where $t$ and $u$ are the usual invariants associated with the
virtual-photon-pomeron sub-process, $F(t_{{\cal P}})$ is a combination of
the Dirac elastic form factor of the proton and the quark \cite{dl}, and $t_{%
{\cal P}}$ is the four-momentum squared of the virtual pomeron. Finally,
integrating over the quark scattering angle and $t_{{\cal P}}$ gives
\begin{equation}
F_2^{D(3)}(\beta ,Q^2,x_{{\cal P}})\propto \beta (1-\beta )(\frac{Q^2}%
\beta )^\lambda
\end{equation}
where we have taken
\begin{equation}
f(t)^2(\frac{-t}{\Lambda ^2})^{2(\alpha _{{\cal P}}-1)}\equiv
 (\frac{t}{\Lambda^2})^{\lambda-1}
\end{equation}
and $\alpha _{{\cal P}}\approx \alpha _{{\cal P}}(0)$. The most obvious
change in the predicted form for the structure function, compared to the
case where the partons are constrained to lie on mass-shell, is the
appearance of a term potentially violating the scale invariance of the cross
 section.
The origin of this term is immediate: since the energy of the quark-proton
diffractive process is
\begin{equation}
s_{diff}=-\frac{k^2}{x_{{\cal P}}}=-\frac{Q^2}{\beta x_{{\cal P}}}(1-\cos
\theta _{cm})  \label{sdiff}
\end{equation}
there is a contribution to the diffractive sub-process amplitude
proportional to
\begin{equation}
(\frac{Q^2}{\Lambda ^2\beta }(1-\cos \theta _{cm}))^{1-\alpha _{{\cal
 P}}}f ^2(%
\frac{Q^2}{\Lambda ^2\beta }(1-\cos \theta _{cm}))  \label{contn}
\end{equation}
If, as is suggested by our analysis of the ladder graphs, $k^2f^2(k^2)$
 is
relatively slowly varying, we expect
\begin{equation}
F_2\propto (\frac{Q^2}{\Lambda ^2\beta })^\lambda
\end{equation}
where
\begin{equation}
\lambda
=2(\alpha _{{\cal P}}-1)\,\hbox{if}\,f(k^2)^2=\frac{1}{k^2}
 \label{scalev}
\end{equation}

Let us consider whether eq(\ref{eq:f2}) can describe the measured events. The
overall power-law behaviour $(x_{{\cal P}})^{2\alpha _{{\cal P}}-1}$ follows
from our Regge parametrisation of the amplitude for the diffractive
sub-process, and is consistent with the experimental measurements, although
ZEUS and H1 find somewhat different values of the exponent \cite{H1,Z2,Z3}:
\bea
\alpha _{\cal P} = 1.09  \pm 0.03 \pm 0.04 \;\; H1 \nonumber \\
\alpha _{\cal P} = 1.15 \pm .04 \pm 0.04 (0.07) \;\;ZEUS2  \nonumber \\
\alpha _{\cal P} =1.23 \pm0.02 \pm 0.04 \;\; ZEUS3
\label{eq:ap}
\eea

As may be seen from Table 1, different pseudo-rapidity cuts give rise to
different constraints on $k_{\min }^2.$ The new Zeus method for extracting
the diffractive contribution does away with the need for such cuts at all
and so will have no constraint on $k_{\min }^2.$ Thus the various
experimental methods probe different distributions of the virtuality of the
struck quark. Given the form of our expression for the contribution of the
graph of Fig.~\ref{fig:1} it is possible that these differences may explain
some of the discrepancies in the results
found such as those in (\ref{eq:ap}). However,
it remains to be seen whether part of the apparent differences in (%
\ref{eq:ap}) could be associated with the different types of event selection.

What about the $\beta$ dependence at fixed $Q^2$? If we ignore the possible
scaling violations, i.e., choose $\lambda = 1$, we predict a hard
distribution $\propto \beta (1- \beta)$. The observed form of the $\beta$
 distribution may be well described by the form \cite{Z3}
\begin{equation}
(\frac{1}{x_{{\cal P}}})^a b( \beta (1-\beta) + \frac{c}{2} (1-\beta)^2)
\label{eq:nikzak}
\end{equation}
with $a, b$ and $c$ constants: $c\approx 0.57$. Thus, (\ref{eq:f2})
 provides a reasonable description at large $\beta$, though it does
fail to reproduce the rise seen at low $\beta$. The situation is somewhat
ameliorated if one allows for  non-zero values of $\lambda$. Taking $\lambda
= 2(1- \alpha_{{\cal P}})$, as would be appropriate to the choice
 $\tilde{f}(k^2) = 1/k^2$, with $\alpha _ {{\cal P}} $ in the range given by the
 experimental
measurements
 (\ref{eq:ap}), generates an enhancement of low-$\beta$ events, but
this is still below the low-$\beta$ growth observed. Thus, whilst there may
be a component of the form presented above, it seems likely that an
additional component may be needed. We shall consider shortly its possible
 origin.

The case of non-zero $\lambda$ leads to a prediction of scaling violation,
correlated with the $\beta $ dependence just discussed. At present, the
experimental situation is somewhat unclear, since H1 has found an indication
of $Q^2$ dependence, whilst ZEUS does not. The H1 results are consistent
with a growth with $Q^2$ of the structure function that is proportional to $%
\log _{10} Q^2$, with coefficients of proportionality $(0.12 \pm 0.09),
\;(0.15 \pm 0.09 ),\; (0.15 \pm 0.09)$ and $(0.17 \pm 0.15)$ for $\beta =
0.65, 0.375, 0.175 $ and $0.065$ respectively. Interpreting these values in
terms of $\lambda$ gives a value of $\lambda = (0.07 \pm 0.05)$. Although
not significant, this is of the correct sign to be interpreted as due to the
term $\propto (\frac{Q^2}{\beta})^{2( \alpha_{{\cal P}}-1)}$ with $\alpha_{%
{\cal P}} > 1$ as is observed, and with a related enhancement of the
structure function at low $\beta$ as just discussed. However, with the
measured value of $\alpha _{{\cal P}} $, the predicted $\lambda$ will be too
large, unless there is a suppression from the form factor $f(k^2)$ beyond that
 chosen in eq(\ref{scalev}).

\section{Summary and Conclusions}

We have re-analysed the viability of the explanation of the diffractive
events observed at HERA based on the quark diffractive scattering graph of
Fig.~\ref{fig:1}.
We have observed that the large {\it pseudorapidity} cuts favoured by early
HERA analyses force the struck quark to be far off its mass shell. This
means that quark diffractive scattering could explain a significant fraction
of these data only if the struck quark could be far off shell without a
significant suppression of the cross section. An immediate implication is
that diffractive events selected in this way do {\it not } probe the
structure function of the pomeron, at least in the normal sense of measuring
the distribution of ``on-shell" partons within a pomeron ``target". Thus our
intuition based on measurements of the structure functions of the nucleon in
conventional deep-inelastic processes does not directly apply to the
interpretation of the structure of the pomeron as revealed by
large-pseudorapidity-gap events\footnote{%
Note also that the much-discussed problems associated with implementing the
momentum sum rule do not arise in our approach. }%
. Our results are in conflict with the commonly-accepted form of the
diffractive quark contributions calculated in \cite{dl}. We have calculated
the full diagram of Fig.~\ref{fig:1}, allowing for a general form factor
which does not cut the integral for the struck quark off close to mass
shell. The resulting form gives a ``hard" distribution which is modified by
a term in which the $\beta$ dependence is correlated with a scaling
violation. While such a term may be able to explain the measurements of
events at large $\beta$, it fails
to reproduce the rise in the
structure function seen at small $\beta$.

\begin{figure}[htp]
\hglue 6.0cm
\epsfig{figure=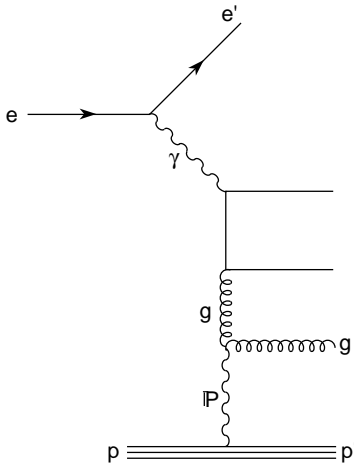,width=4.75cm}
\caption[]{Diffractive gluon component of the pomeron.}
\label{fig:5}

\end{figure}

The question immediately arises: what may the missing contribution be? An
obvious candidate is the graph of Fig.~\ref{fig:5}, in which the pomeron
couples to a gluonic component of the virtual photon. The reason this may be
significant at low $\beta$ is that graphs of this type have an infra-red
singularity when the final-state gluon is soft, which leads to a $1/\beta$
contribution to the amplitude squared. This should be compared to the case
of Fig.~\ref{fig:1}, in which the amplitude squared is constant for small $%
\beta$ because there is no equivalent soft singularity. As a result, the
graph in of Fig.~\ref{fig:5} makes a contribution to the structure function
which is constant at low $\beta$. Given that the pomeron may couple more
strongly to gluons than to quarks, it is plausible that this graph may
generate a significant contribution in the low $\beta$ region. Together with the
 quark contribution this may provide a good description of the form of
(\ref{eq:nikzak}).  This particular form was
motivated by a two-gluon model for the pomeron investigated by Nikolaev and
Zakharov \cite{nz}. However, we see that the essential feature, namely the
absence of a fall-off at low $\beta$, does not rely on a two-gluon
Lipatov-like interpretation for the pomeron, but simply reflects the
characteristic infra-red behaviour associated with gluon emission, such as
in Fig.~\ref{fig:5}.

Thus we arrive at a perfectly consistent picture of the large-$pseudorapidity
$-gap HERA diffractive events, interpreted in terms of the diffractive
scattering of (virtual) partons on the proton. It will be particularly
interesting if the $Q^2$ dependence of the large-$\beta$ events can be
extracted reliably, for there should be some scaling violation associated
with the $(Q^2)^{\lambda} $ factors discussed above.

\end{document}